\documentclass [11pt]{article}
\usepackage {graphicx}

\title{Hyperbolic chaos in self-oscillating systems based on mechanical triple
linkage: Testing absence of tangencies of stable and unstable manifolds for
phase trajectories}

\author{S.P. Kuznetsov $^{1, 2}$}

\begin{document}

\maketitle
\begin{center}
$^{1}$ Institute of Computer Science, Udmurt State University,\newline
Universitetskaya 1, Izhevsk, 426034, Russia

$^{2}$ Kotelnikov's Institute of Radio-Engineering and
Electronics of RAS, Saratov Branch, Zelenaya 38, Saratov, 410019, Russia
\end{center}

\begin{abstract}
\noindent
Dynamical equations are formulated and a numerical study
is provided for self-oscillatory model systems based on the triple linkage
hinge mechanism of Thurston -- Weeks -- Hunt -- MacKay. We consider systems
with holonomic mechanical constraint of three rotators as well as systems, where
three rotators interact by potential forces. We present and discuss some
quantitative characteristics of the chaotic regimes (Lyapunov exponents,
power spectrum). Chaotic dynamics of the models we consider are associated
with hyperbolic attractors, at least, at relatively small supercriticality
of the self-oscillating modes; that follows from numerical analysis of the
distribution for angles of intersection of stable and unstable manifolds of
phase trajectories on the attractors. In systems based on rotators with
interacting potential the hyperbolicity is violated starting from a certain
level of excitation.
\end{abstract}

\bigskip


\section{Introduction}

Hyperbolic theory is a branch of the theory of dynamical systems undergoing
essential and deep development during the last half century. It gives rigorous
mathematical justification of chaotic behavior in deterministic systems,
both in the discrete-time case (iterative maps -- diffeomorphisms) and in
continuous time (flows)~\cite{1,2,3,4,5}. The hyperbolic theory deals with
invariant sets in phase space of dynamical systems composed exclusively
of saddle trajectories. For such a trajectory, in the vector space of all
possible infinitesimal perturbations (tangent space), one can define a
subspace of vectors of exponentially decreasing norm in direct time, and a
subspace of vectors exponentially decreasing in reverse time. In flow
systems, for trajectories different from a fixed point, one has to introduce
additionally a neutral one-dimensional subspace corresponding to
perturbations along the phase trajectory, which neither grow, nor decay in
time in average. An arbitrary vector of small perturbation is required to be
a linear combination of vectors related to these subspaces. The set of
trajectories that approach the reference orbit in the course of evolution in
time is called the stable manifold. Similarly, the unstable manifold is a
set of trajectories which approach the reference orbit in reverse time.

For conservative systems the hyperbolic chaos is associated with Anosov
dynamics, when a typical trajectory in the phase space (for
diffeomorphisms), or on the energy surface (for flows), is dense. For
dissipative systems, the hyperbolic theory introduces a special type of
chaotic attractors called the uniformly hyperbolic attractors.

The dynamic behavior considered by the hyperbolic theory is rough, or
structurally stable, that is, the phase space arrangement, dynamical
behavior and its statistical characteristics are not sensitive to variations
in parameters and functions in the equations of motion. In this regard, it
was natural to expect that the hyperbolic chaos should occur in many
physical situations. However, as time passed and numerous examples of
chaotic systems of different nature were proposed and studied, it became
clear that they do not fit the narrow frames of the early hyperbolic theory.
Therefore, the hyperbolic dynamics is regarded commonly rather as a kind of
refined abstract image of chaos than something having direct relation to
real systems. So, efforts of mathematicians were redirected to development
of broadly applicable generalizations ~\cite{6,7}.

In textbooks and reviews on dynamical systems the hyperbolic chaos is
represented usually by artificially constructed mathematical examples, like
Anosov torus automorphism, DA-attractor of Smale, Smale -- Williams
solenoid, Plykin attractor ~\cite{1,2,3,4,5}, whereas the question of implementation and
possible applications of hyperbolic chaos in nature and technology was not
elaborated for a long time.

If mathematicians are developing their examples using geometric,
topological, algebraic constructions, a physicist should use a very
different toolbox for design models with hyperbolic chaos: oscillators,
particles, interactions, feedback loops etc. Recently, great progress has
been made in this direction, and numerous examples of physically realizable
systems are offered with attractors of Smale -- Williams type and with
other kinds of hyperbolic attractors~\cite{8,9,10,11,12,13,14,15,16,17,18,19,20}. Regarding clarity and
transparency, a preference should be given surely to mechanical systems~\cite{20,21,22,23,24} as they are easily understood and interpreted through our everyday
experience.

It may be noted, however, that the physical examples of hyperbolic
attractors discussed so far~\cite{8,9,10,11,12,13,14,15,16,17,18,19,20} are obtained by reduction of the
dynamical description to the Poincar\'{e} maps; in time intervals between
the passages of the Poincar\'{e} section, one cannot speak definitely of
uniform in time stretching and compression for the corresponding
vector subspaces of perturbations. The question of design of physical systems with
attractors, which would be characterized by hyperbolicity uniform in
continuous time, at least approximately, remains open.

In this regard, an interesting starting point example is the hinge mechanism
discussed in the popular-scientific article of Thurston and Weeks~\cite{25} as
an illustration of a system with nontrivial topology of the configuration
space. The mechanism is composed of three identical disks in a common
plane, which are able to rotate about their central axes fixed at the
vertices of an equilateral triangle (Fig. 1). At the edge of each disk a
hinge is attached, and these hinges P$_{1,2,3}$ are connected by three
identical rods to one more movable hinge P$_{0}$.

Instant configuration of the system is determined by the variables $\,\theta
_1 ,\,\theta _2 ,\,\theta _3 $ characterizing the rotation angles of the
disks, but only two of them are independent because of the imposed
constraint. Thus, the configuration space is a two-dimensional manifold. In
settings we consider here it will be a surface of genus 3 ("pretzel with
three holes"). The kinetic energy is a quadratic form of the generalized
velocities (time derivatives of the local coordinates on the
two-dimensional manifold). This quadratic form whose coefficients depend on
lengths and masses of the structural elements, defines a certain metric on the
two-dimensional manifold, and the motion takes place along the geodesic
lines of this metric.

It was proven that for two-dimensional manifolds
of genus different from 0 and 1 the dynamics along the geodesics
are nonintegrable~\cite{25a}.
If the curvature is negative everywhere, the motion definitely corresponds to Anosov
hyperbolic dynamics~\cite{27,28}. In the triple linkage, as stated by Hunt and MacKay~\cite{26},
with appropriate
selection of sizes and masses
one can reach a situation where the metric is of negative
curvature over the whole manifold. Recently, some other hinge mechanisms
capable of demonstrating the dynamics of Anosov have been
proposed and analyzed~\cite{28a,28b}.
(Moreover, similar dynamics are discussed in
the context of model description of motion of electrons in the
doubly periodic potential of two-dimensional crystal lattice~\cite{28c,26}.)

Hunt and MacKay also pointed out~\cite{26} that with
addition of friction and providing feedback by means of a supplied control
device one can get a hyperbolic chaotic attractor. However, in this respect,
the authors limit themselves to though convincing, but only to verbal
argumentation, appealing to the structural stability of the Anosov dynamics
in the original conservative system. No study was carried out which would
consider concrete equations of a model system, demonstrate hyperbolic chaos
numerically, and analyze its quantitative characteristics like Lyapunov
exponents.

\begin{center}
\begin{figure}[htbp]
\centerline{\includegraphics[width=4in]{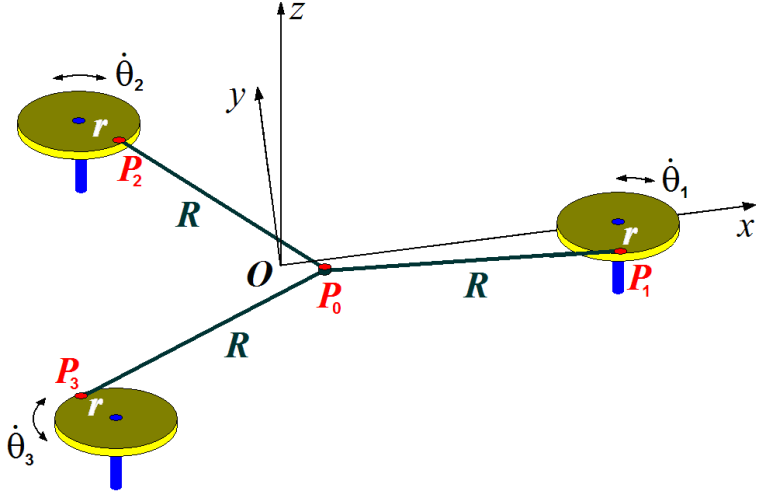}}
\label{fig1}
\caption{The triple linkage of Thurston and Weeks.}
\end{figure}
\end{center}

In the present article we formulate differential equations and provide a
numerical study of self-oscillatory systems inspired by the triple linkage
mechanism. The analysis here will be restricted to the case where the disk
radius is small compared with the length of the rods connecting them with a
movable hinge, and inertial properties of the mechanism are determined
exclusively by the disks. In this case, the curvature of the metric defined
by the kinetic energy is negative except for a finite set of eight points
where it is zero. It is definitely enough to ensure the Anosov hyperbolic
dynamics at any positive constant kinetic energy.

First, we will discuss a model precisely corresponding to the idea of
Hunt and MacKay, when dissipation is introduced depending on the total
kinetic energy of the system, being negative at small energies and positive at large energies;
it creates a situation where some constant-energy surface becomes an
attractor. Practically it implies that the added control device has to
measure instantaneous kinetic energy of the whole system and actuators have
to be introduced which provide application of torques to the disks to make
the energy approach a desirable fixed value.

Another model system we examine is based on the assumption that each disk is a
self-rotating subsystem; without interaction with the partners it would tend
to rotation in one or other direction with some constant angular velocity
not depending on specific initial conditions, due to the dissipation and
feedback intrinsic to this subsystem itself. This variant seems much easier
for practical implementation; it is possible to use a pair of friction
clutches attached to each disk and passing rotation in opposite directions,
with a properly chosen law of friction dependence on the relative angular
velocity. When mechanical constraints are imposed by rods and hinges in such a
device, the dynamics will not be reduced to that on a constant energy
surface; the kinetic energy after decay of transients continues to oscillate
chaotically.

Next, instead of the models with geometric constraints, we consider systems
of three rotators interacting through some force field whose potential
depends on the angles of rotation of the disks (see also Ref.~\cite{29}). The
potential minimum is supposed to take place on a surface in the
configuration space assumed in the previous models with mechanical
constraint. It basically allows us to get away from the mechanical objects
and to talk about possible implementation of chaotic systems of other
physical nature, like electronic devices operating as generators of robust
chaos.

As we depart from the original hinge mechanism, where chaos is hyperbolic,
the hyperbolicity surely persists due to the structural stability of the
Anosov dynamics for infinitesimal variations of the evolution operator in
comparison with the original system. But for variations of finite magnitude,
justification of the hyperbolic nature of the dynamics is of 
significance; so we have to apply some computer criteria for verification of
hyperbolicity. In the present paper we will exploit the so-called criterion
of angles for this purpose.

The idea of testing hyperbolicity based on statistics of angles between
stable and unstable manifolds has been proposed for saddle invariant sets in~\cite{30}. Subsequently, it has been used for verification of hyperbolicity of
attractors~\cite{31,32}, including attractors of Smale -- Williams and Plykin
type in Poincar\'{e} maps for systems that allow physical implementation~\cite{8,13,33}.

The technique consists in the following. Along a typical phase trajectory
belonging to the invariant set of interest, we follow it in forward time and
in reverse time, and evaluate at each point an angle between the subspaces
of perturbation vectors to analyze their statistical distribution. If the
resulting distribution does not contain angles close to zero, it indicates a
hyperbolic nature of the invariant set. If positive probability for zero
angles is found, the tangencies between the stable and unstable manifolds of
the trajectory do occur, and the invariant set is not hyperbolic. The latter
may indicate the presence of a quasi-attractor that is a complex set, which
contains long-period stable cycles with very narrow domains of attraction~\cite{2}.

\section{A system with a mechanical constraint \\ possessing an invariant energy surface}

The Cartesian coordinates of the hinges P$_{1,2,3}$ attached to the disks
(Figure 1) are expressed through the angles $\theta _{1}$, $\theta
_{2}$, $\theta _{3}$, measured from the rays connecting the centers of
the disks with origin O, as follows:
\begin{equation}
\label{eq1}
\begin{array}{l}
x_1=1-r\cos\theta_1,\,\,y_1=-r\sin\theta_1,\,\,\\
x_2=-\textstyle{1\over2}+\textstyle{1\over2}r\cos\theta_2+
\textstyle{{\sqrt3}\over2}r\sin\theta_2,\,\,\,\,y_2=
\,\textstyle{{\sqrt3}\over2}+\textstyle{1\over2}r\sin\theta_2-
\textstyle{{\sqrt3}\over2}r\cos\theta_2,\,\,\\
x_3=-\textstyle{1\over2}+\textstyle{1\over2}r\cos\theta_3-
\textstyle{{\sqrt3}\over2}r\sin\theta_3,\,\,\,\,\,y_3=-
\textstyle{{\sqrt3}\over2}+\textstyle{1\over2}r\sin\theta_3+
\textstyle{{\sqrt3}\over2}r\cos\theta_3.\\
\end{array}
\end{equation}

The mechanical geometric constraint due to the rods and hinges implies that the
radius of the circumscribed circle of the triangle P$_{1}$P$_{2}$P$_{3}$
must be $R$. By the well-known formula, $R = abc / 4S$, where $a$, $b$, $c$ are the
lengths of sides of the triangle, and $S$ is its area. With given coordinates
of three vertices ($x_{i}$, $y_{i})$, we evaluate the lengths of sides, and
the area is expressed through the vector product of two vectors, one from
P$_{1}$ to P$_{2}$ and the other from P$_{1}$ to P$_{3}$: $S = \textstyle{1
\over 2}\left| {{\rm {\bf b}}\times {\rm {\bf c}}} \right|$. Thus, we have
$(abc)^2 - 4R^2({\rm {\bf b}}\times {\rm {\bf c}})^2 = 0$, which can be
rewritten as
\begin{equation}
\label{eq2}
\begin{array}{l}
[(x_1-x_2)^2+(y_1-y_2)^2][(x_2-x_3)^2+(y_2-y_3)^2][(x_3-
x_1)^2+(y_3-y_1)^2]-\\
\,\,\,\,\,\,\,\,\,\,\,\,\,\,\,\,\,\,\,\,\,\,\,-4R^2(x_2y_3+x_3y_1+
x_1y_2-x_3y_2-x_1y_3-x_2y_1)^2=0.\\
\end{array}
\end{equation}
Substituting the expressions (\ref{eq1}), we get the geometric constraint equation in the form
$F(\theta_1 ,\theta_2 ,\theta_3 )=0$.

Let us suppose that the disks are the only massive elements of the
construction, and their moments of inertia are assumed to be unity;
additionally, each of them may be driven by an external torque
$M_{1,2,3}$. Then the equations of motion read (see, e.g.,~\cite{34,35})
\begin{equation}
\label{eq3}
\ddot{\theta}_1=M_1+\Lambda\partial F/\partial\theta_1,\,\,\ddot
{\theta}_2=M_1+\Lambda\partial F/\partial\theta_2,\,\,\ddot
{\theta}_3=M_3+\Lambda\partial F/\partial\theta_3,
\end{equation}
\begin{equation}
\label{eq4}
F(\theta_1,\,\,\theta_2,\,\,\theta_3)=0,
\end{equation}
where the multiplier $\Lambda $ must be determined taking into account the
algebraic condition of the holonomic mechanical constraint (\ref{eq4}). This
expression $F = 0$ and the relation obtained by differentiating
$\sum\nolimits_{j=1}^3{\dot{\theta}_j}\partial F/\partial\theta_j=0$ correspond to two integrals of motion for the system (\ref{eq3}), which is
formally of the sixth order.

Assuming $r<<1$ and expanding (\ref{eq2}) in Taylor series up to terms of the
first order in the small parameter, we obtain
\begin{equation}
\label{eq5}
27(R^2-1)-18r(2R^2-3)(\cos\theta_1+\cos\theta_2+\cos\theta_3)=0.
\end{equation}
Setting $R$=1,we arrive at a very simple equation of holonomic constraint
\begin{equation}
\label{eq6}
F(\theta_1,\theta_2,\theta_3)=\cos\theta_1+\cos\theta_2+\cos
\theta_3=0,
\end{equation}
which will be the only one used in subsequent considerations.

In the conservative case, in the absence of the external torques,
$M_{1,2,3}$=0, the system conserves the kinetic energy
\begin{equation}
\label{eq7}
T=\textstyle{1\over2}(\dot{\theta}_1^2+\dot{\theta}_2^2+\dot
{\theta}_3^2),
\end{equation}
and the dynamics can be interpreted as motion of a point particle on a
two-dimensional surface given by equation (\ref{eq6}), along the geodesics of the metric defined by the quadratic form
\begin{equation}
\label{eq8}
ds^2=\textstyle{1\over2}(d\theta_1^2+d\theta_2^2+d\theta_3^2),
\end{equation}
with the consistence condition on the differentials
$d\theta_1\sin\theta_1+d\theta_2\sin\theta_2+d\theta_3\sin\theta_3=0$ because of
the constraint equation. Calculation of the Gaussian curvature leads to the
formula~\cite{26,29}
\begin{equation}
\label{eq9}
K=-\frac{\cos^2\theta_1+\cos^2\theta_2+\cos^2\theta_3}{2\left(
{\sin^2\theta_1+\sin^2\theta_2+\sin^2\theta_3}\right)^2}.
\end{equation}
As is seen, the curvature is negative everywhere, except for a finite number of
points, namely, eight of them, where it vanishes:
$(\theta_1,\,\theta_2,\,\theta_3)=(\pm\pi/2,\,\pm\pi/2,\,\pm\pi/2)$.
Therefore, the
motion along geodesic lines with nonzero kinetic energy is the Anosov
dynamics~\cite{26}.

Formally, the phase space of the system is six-dimensional; accordingly,
there are six Lyapunov exponents characterizing the behavior of phase
trajectories perturbed near a reference orbit. In the conservative case,
there is one positive, four zero, and one negative exponent, which is equal
to the positive exponent in absolute value. One exponent equal to zero
appears due to the autonomous nature of the system; it is responsible for a
perturbation vector tangent to the reference phase trajectory. One more zero
exponent is associated with a perturbation corresponding to the energy
shift. The remaining two zero exponents are nonphysical and should be
excluded from consideration, since they relate to perturbations of two
integrals of motion caused by the mechanical constraint, i.e., they violate
the constraint equation.

Since the system does not have any intrinsic characteristic time scale, the
positive and negative Lyapunov exponents for the exponential growth and
decay of perturbations per unit time should be proportional to the
velocity, i.e., to the square root of energy, namely, $\lambda = \pm \kappa
\sqrt W $. As the dynamics is associated with motion on the surface of
negative curvature, the coefficient in this expression is determined by the
average curvature of the metric. Empirical numerical calculations for the
system with the constraint (\ref{eq6}) yield $\kappa = 0.70$~\cite{29}.

To get a dissipative system with chaotic attractor according to the
proposals of Hunt and MacKay~\cite{26}, we introduce the torques depending on the
generalized velocities in such a way that the kinetic energy tends during
the time evolution to a value $\mu $, namely
\begin{equation}
\label{eq10}
M_i=\nu[\mu-\textstyle{1\over2}(\dot{\theta}_1^2+\dot{\theta
}_2^2+\dot{\theta}_3^2)]\dot{\theta}_i,\,\,i=1,2,3,
\end{equation}
where $\nu$ is a constant factor. To obtain such a function, the mechanism
should be supplemented with a controlling device and actuators, which apply
the torques $M_{1,2,3}$ to the axles of the disks depending on the value of
the detected instant kinetic energy.

Differentiating the constraint equation with respect to time once and twice, we
have
\begin{equation}
\label{eq11}
\begin{array}{c}
\dot{\theta}_1\sin\theta_1+\dot{\theta}_2\sin\theta_2+\dot{\theta
}_3\sin\theta_3=0,\\
\ddot{\theta}_1\sin\theta_1+\ddot{\theta}_2\sin\theta_2+\ddot
{\theta}_3\sin\theta_3+\dot{\theta}_1^2\cos\theta_1+\dot
{\theta}_2^2\cos\theta_2+\dot{\theta}_3^2\cos\theta_3=0.\\
\end{array}
\end{equation}
Substituting the second derivatives from the equations of motion (\ref{eq3}) into (\ref{eq11}), we obtain an explicit formula for the multiplier
\begin{equation}
\label{eq12}
\Lambda = - \frac{\sum\nolimits_{j = 1}^3 {\left( {\dot {\theta }_j^2 \cos
\theta _j + M_j \sin \theta _j } \right)} }{\sum\nolimits_{j = 1}^3 {\sin
^2\theta _j } },
\end{equation}
and arrive at the closed set of equations
\begin{equation}
\label{eq13}
\ddot {\theta }_i = \nu [\mu - \textstyle{1 \over 2}(\dot {\theta }_1^2 +
\dot {\theta }_2^2 + \dot {\theta }_3^2 )]\left( {\dot {\theta }_i -
\frac{\sum\nolimits_{j = 1}^3 {\dot {\theta }_j \sin \theta _j }
}{\sum\nolimits_{j = 1}^3 {\sin ^2\theta _j } }\sin \theta _i } \right) -
\frac{\sum\nolimits_{j = 1}^3 {\dot {\theta }_j^2 \cos \theta _j }
}{\sum\nolimits_{j = 1}^3 {\sin ^2\theta _j } }\sin \theta _i ,\,\,i =
1,2,3.
\end{equation}

The equation determining the time evolution of the kinetic energy 
$W =\textstyle{1 \over 2}(\dot {\theta }_1^2 + \dot {\theta }_2^2 + \dot {\theta}_3^2 )$ 
is derived easily from (\ref{eq13}) and reads
\begin{equation}
\label{eq14}
\dot {W} = 2\nu (\mu - W)W.
\end{equation}
\begin{figure}[htbp]
\centerline{\includegraphics[width=5.5in]{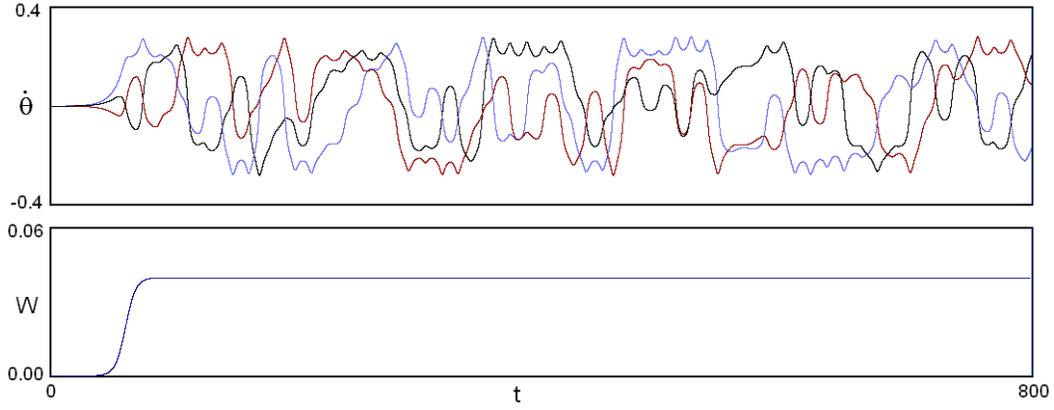}}
\label{fig2}
\caption{
Dependences of the generalized velocities
$\dot {\theta }_1,\,\,\dot {\theta }_2 ,\,\,\dot {\theta }_3 $ and the energy $W$ on time in the
course of the transient process from a low-energy state towards chaotic
self-oscillations in the system (\ref{eq13}) at $\mu=0.04 $, $\nu=3$.
}
\end{figure}

Figure 2 illustrates the transient process in the system (\ref{eq13}) at $\mu $=0.04
and $\nu $=3 starting from a point in the configuration space consistent
with the mechanical constraint with very small initial velocity towards the
regime of chaotic self-oscillations, which corresponds to constant energy,
although the angular velocities behave chaotically, without any visible
repetition of forms.

Figure 3 shows a trajectory in the configuration space obtained from
numerical integration of equations (\ref{eq13}). Due to the imposed mechanical
constraint, it is placed on a two-dimensional surface defined by the equation
$\cos \theta _1 + \cos \theta _2 + \cos \theta _3 = 0$. When plotting the
picture, the angular variables are related to the interval from 0 to 2$\pi$, 
i.e., the diagram in the three-dimensional space ($\theta _{1}$,$\theta_{2}$,$\theta _{3})$ 
corresponds to a single fundamental cubic cell,
which reproduces itself periodically with a shift by 2$\pi $ along each of
three coordinate axes. Any opposite pair of faces of the cubic cell may be
identified; then we come to a compact manifold of genus 3, i.e., to the
surface topologically equivalent to the "pretzel with three holes."

\begin{figure}[htbp]
\centerline{\includegraphics[width=2.3in]{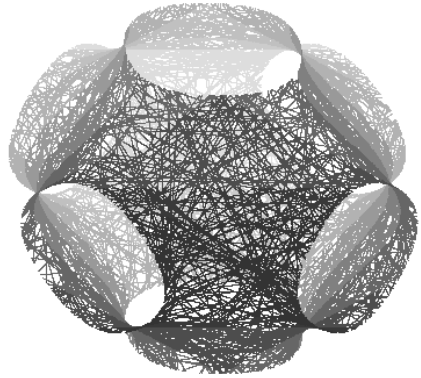}}
\label{fig3}
\caption{ A trajectory in three-dimensional configuration space of the system (\ref{eq13}) obtained by numerical integration of the equations; it is placed on the two-dimensional surface defined by the constraint equation $\cos \theta _1 + \cos \theta _2 + \cos \theta _3 = 0$.}
\end{figure}

In order to characterize the observed chaos quantitatively, we turn to the
Lyapunov exponents. Excluding two nonphysical zero exponents, which violate
the constraint equation, we have four exponents in the rest. Since the
motion on the attractor takes place on the energy surface, the exponents for
perturbations without departure from this surface will be equal to those in
the conservative system at the same energy: $\kappa \sqrt W $, 0, and $ -
\kappa \sqrt W $. The exponent corresponding to perturbation of the energy
is evaluated simply as the Lyapunov exponent of the attracting fixed point
$W = \mu $ in the equation (\ref{eq14}); it is equal to $ - 2\nu \mu $.

Figure 4 shows graphs for the Lyapunov exponents versus parameter $\mu $ in
the system (\ref{eq13}). The solid lines correspond to the relations mentioned in the preceding paragraph. The dots represent results of calculation of the
Lyapunov exponents using the Benettin algorithm~\cite{36,13}. Numerical
integration of equations (\ref{eq13}), formally written in the form ${\rm {\bf
\dot {x}}} = {\rm {\bf F}}({\rm {\bf x}},t)$, where \textbf{x} is a
six-dimensional state vector, is performed together with a collection of six
sets of variation equations ${\rm {\bf \dot {\tilde {x}}}} = {\rm {\bf
{F}'}}({\rm {\bf x}}(t),t){\rm {\bf \tilde {x}}}$, where ${\rm {\bf
{F}'}}({\rm {\bf x}}(t),t)$ is the Jacobi matrix of size 6x6 composed of 
partial derivatives. In the process of numerical integration,
orthogonalization and normalization of the perturbation vectors is carried
out at each step, according to the Gram -- Schmidt procedure. Lyapunov
exponents are evaluated as coefficients of increase or decrease of the
accumulated sums of logarithms of vector norms obtained after the
orthogonalization, but before the normalization. At the final stage of the
procedure, two nonphysical exponents violating the mechanical constraint
condition are excluded.
\begin{figure}[htbp]
\centerline{\includegraphics[width=3.5in]{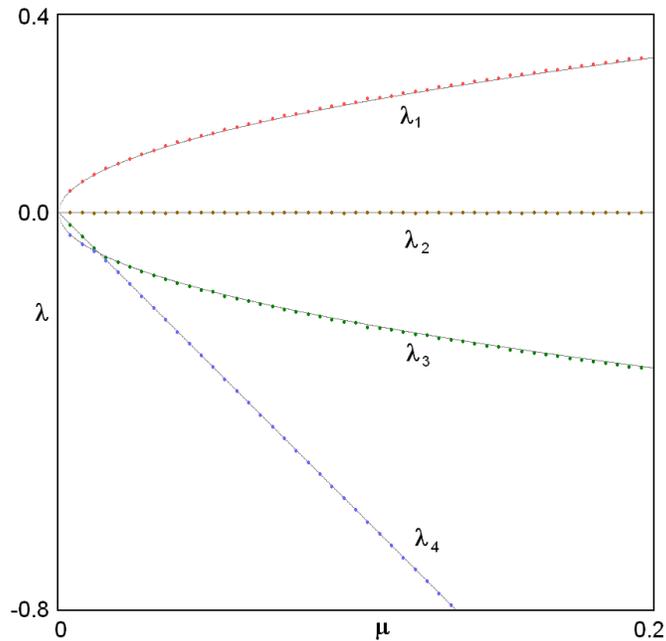}}
\label{fig4}
\caption{ Lyapunov exponents for the dissipative system (\ref{eq13}) depending on the parameter $\mu $ at $\nu $=3. The dots correspond to numerical
computations based on the Benettin algorithm, and the solid lines are
plotted according to the formulas $\pm 0.70\sqrt \mu $ and $ - 2\nu \mu $.}
\end{figure}
\begin{center}
\begin{figure}[htbp]
\centerline{\includegraphics[width=3in]{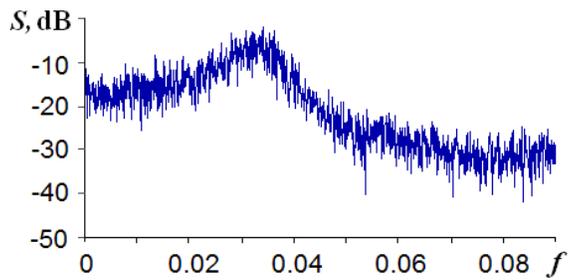}}
\label{fig5}
\caption{ The power spectrum of the signal $u(t) = \cos \theta _1 (t)$ from
the system (\ref{eq13}) for $\nu $ = 3, $\mu $ = 0.04}
\end{figure}
\end{center}
\begin{figure}[htbp]
\centerline{\includegraphics[width=5.5in]{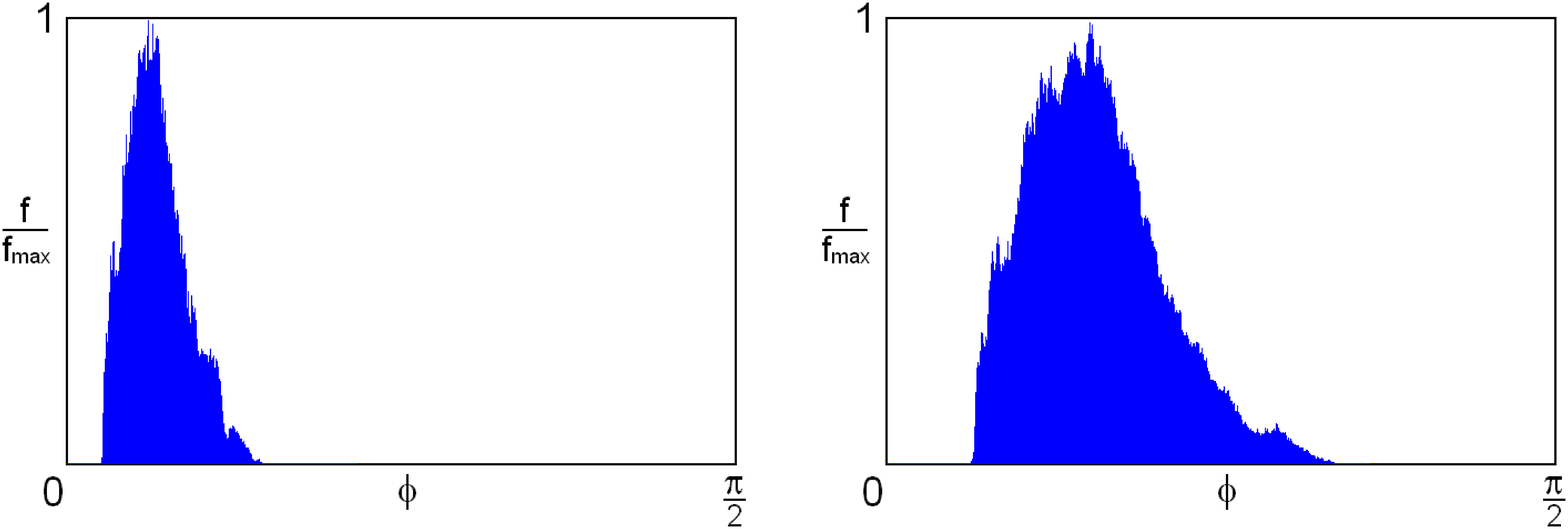}}
\label{fig6}
\caption{Verification of the criterion of angles in the system (\ref{eq13}) with a mechanical constraint for the attractor represented by an invariant energy
surface at $\mu $=0.04 (a) and at $\mu $=0.25 (b) with $\nu $=3.}
\end{figure}

Figure 5 shows the power spectrum of the signal determined as $u(t) = \cos
\theta _1 (t)$. It is obtained by processing time series from numerical
simulation of the dynamics on the attractor at $\mu $ = 0.04 and $\nu $ = 3.
The spectrum is computed in accordance with the procedure of the power
spectral density estimation recommended in the theory of stochastic
processes~\cite{37}. The spectral density is depicted in logarithmic scale, in
decibels (10~dB correspond to the 10-fold ratio of power levels). As seen
from the diagram, the continuous spectrum, which corresponds to chaotic
dynamics, is quite uniform (no pronounced peaks are seen).

The observed attractor is undoubtedly hyperbolic since the dynamics takes
place along geodesic lines of the metric of negative curvature (except a
finite number of points) on the energy surface. It is interesting, however,
to try application of the criterion of angles in this case to test the
methodology, which further will be exploited in situations where the
presence or absence of hyperbolicity is not trivial.

To use the criterion in a form proposed in~\cite{33}, we
start by calculating a reference orbit ${\rm {\bf x}}(t)$ on the
attractor, tracing the solution of the equation ${\rm {\bf \dot {x}}} = {\rm
{\bf F}}({\rm {\bf x}},t)$ over a sufficiently long time interval. Then take
the linearized equation for the perturbation vector ${\rm {\bf \dot {\tilde
{x}}}} = {\rm {\bf {F}'}}({\rm {\bf x}}(t),t){\rm {\bf \tilde {x}}}$ and
integrate it along the reference trajectory forward in time with
normalization of the vector ${\rm {\bf \tilde {x}}}$ at each step $n$ to avoid
divergence. (In our problem we have one unstable direction, and, so, one
positive Lyapunov exponent.) The result is a set of unit vectors $\,\{{\rm
{\bf x}}_n \}$. Next, the integration is performed in reverse time along the
same reference orbit using the linear equation ${\rm {\bf \dot {u}}} = -
[{\rm {\bf {F}'}}({\rm {\bf x}}(t),t)]^{\rm{T}}{\rm {\bf u}}$, where the
superscript T denotes the matrix conjugation. It provides a set of vectors
$\,\{{\rm {\bf u}}_n \}$, each defines the orthogonal complement to the
stable subspace of the perturbation vectors at a point of the reference
trajectory. These vectors also are normalized to unity in the course of the
computations. Now, to evaluate the angle $\phi $ between the one-dimensional
unstable subspace and the stable subspace at each $n$-th step we calculate the
angle $\beta _{n} \in $[0,$\pi $/2] between the vectors ${\rm {\bf
\tilde {x}}}_n $ and ${\rm {\bf u}}_n $: $\cos \beta _n = \left| {{\rm {\bf
u}}_n (t) \cdot {\rm {\bf \tilde {x}}}_n (t)} \right|$, and set $\phi _n =
\pi \mathord{\left/ {\vphantom {\pi 2}} \right. \kern-\nulldelimiterspace} 2
- \beta _n $.

Figure 6 shows the histograms obtained numerically for the angles between
stable and unstable subspaces in the system (\ref{eq13}). As is seen, the distributions
are well separated from zero angles $\phi $, i.e., the test confirms the
hyperbolic nature of the attractor.

\section{A system of three self-rotators \\ with a mechanical constraint}

Let us turn now to a system where the external forces applied to the disks
depend only on the angular velocity of each disk, namely, suppose
\begin{equation}
\label{eq15}
M_i = \nu (\mu - \textstyle{1 \over 2}\dot {\theta }_1^2 )\dot {\theta }_i
,\,\,i = 1,2,3.
\end{equation}

In a real device, to provide such torques, one can use a pair of friction
clutches attached to each disk, which transmit oppositely directed
rotations, selecting properly the functional dependence of the friction
coefficient on the velocity. In this case, each of the three disks is a
self-rotator that means a subsystem, which, being singled out, manifests
evolution in time with approach to a steady rotation with constant angular
velocity $\dot {\theta } = \pm \sqrt {2\mu } $ in one or other direction
(depending on initial conditions).

Equations (\ref{eq3}) upon the constraint (\ref{eq4}) in this case can be rewritten as
\begin{equation}
\label{eq16}
\ddot {\theta }_i = \nu (\mu - \textstyle{1 \over 2}\dot {\theta }^2)\left(
{\dot {\theta }_i - \frac{\sum\nolimits_{j = 1}^3 {\dot {\theta }_j \sin
\theta _j } }{\sum\nolimits_{j = 1}^3 {\sin ^2\theta _j } }\sin \theta _i }
\right) - \frac{\sum\nolimits_{j = 1}^3 {\dot {\theta }_j^2 \cos \theta _j }
}{\sum\nolimits_{j = 1}^3 {\sin ^2\theta _j } }\sin \theta _i ,\,\,i =
1,2,3.
\end{equation}

During the motion, due to the imposed mechanical constraint, the
representative point remains on the same surface as in the previous example
(Figure 3).

Figure 7 shows time dependences of the angular velocities of the disks and
of the energy $W = \textstyle{1 \over 2}(\dot {\theta }_1^2 + \dot {\theta
}_2^2 + \dot {\theta }_3^2 )$ in the transient process at $\mu $=0.02, $\nu
$=3 starting from a point allowable by the mechanical constraint with some
small initial velocity. As a result, a regime of chaotic self-oscillations
develops, in which the kinetic energy undulates irregularly around a mean
value. Figure 8 shows how the kinetic energy and standard deviation depend
on the super-criticality parameter $\mu $.
\begin{figure}[htbp]
\centerline{\includegraphics[width=5.5in]{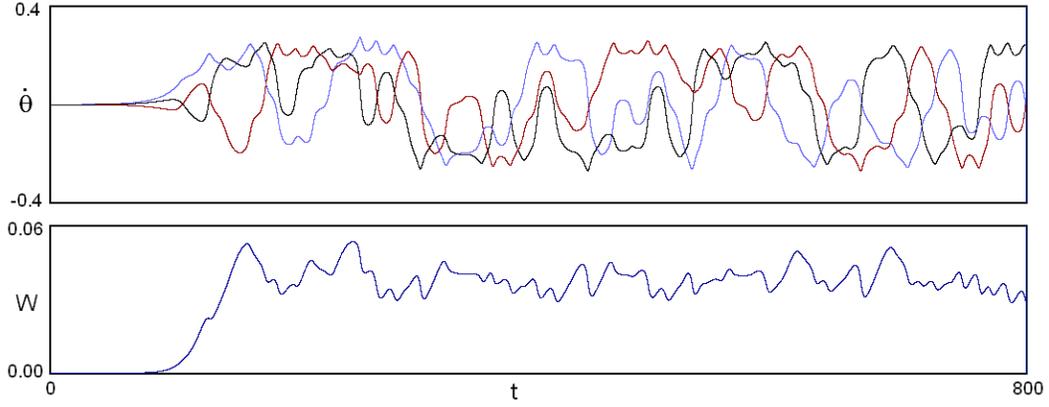}}
\label{fig7}
\caption{Dependences of the generalized velocities $\dot {\theta }_1
,\,\,\dot {\theta }_2 ,\,\,\dot {\theta }_3 $ and energy $W = \textstyle{1
\over 2}(\dot {\theta }_1^2 + \dot {\theta }_2^2 + \dot {\theta }_3^2 )$ on
time in the transient process of development of chaotic self-oscillations in
the mechanical linkage of three self-rotators (\ref{eq16}) with $\mu $=0.02, $\nu $=3.}
\end{figure}

Concerning a number of Lyapunov exponents for the new version of
the system, the same arguments as for the previous model are valid; so, four
of them are relevant. Figure 9 shows a plot of the Lyapunov exponents versus
parameter $\mu $ as computed with the Benettin algorithm. The diagram looks
qualitatively similar to that in Fig. 4; so, one can suppose that the
hyperbolicity persists despite of the occurrence of energy oscillations in
time.

Figure 10 shows the power spectrum of the signal $u(t) = \cos \theta _1 (t)$
as obtained by processing the time series from numerical simulation of the
dynamics on the attractor that demonstrates obvious similarity with Figure
5. Like in the previous case, the spectrum is continuous and looks uniform
enough.

\begin{figure}[htbp]
\centerline{\includegraphics[width=2.5in]{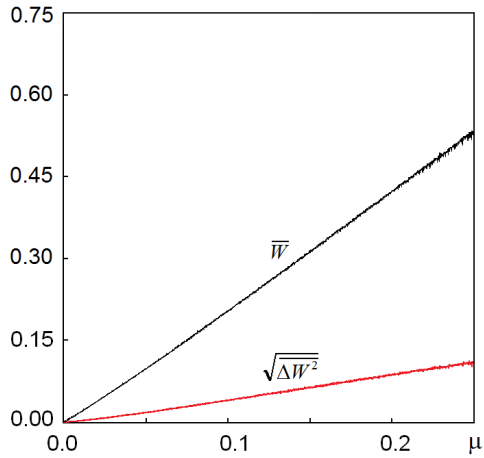}}
\label{fig8}
\caption{Dependences of the kinetic energy and its standard deviation on
the parameter $\mu $ as obtained from numerical integration of equations
(\ref{eq16}) for the mechanical linkage of three self-rotators at $\nu $=3.}
\end{figure}

\begin{figure}[htbp]
\centerline{\includegraphics[width=3.5in]{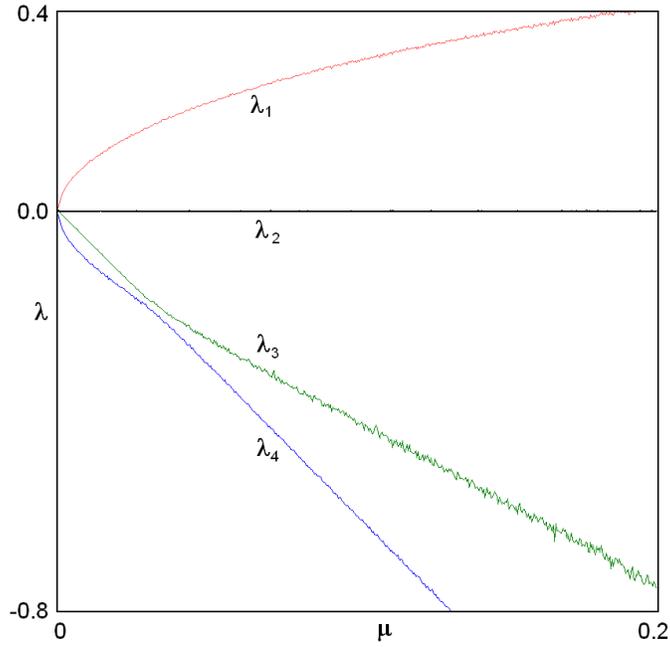}}
\label{fig9}
\caption{ Lyapunov exponents of the linkage of three self-rotators (\ref{eq16})
depending on $\mu $ at $\nu $=3.}
\end{figure}

\begin{center}
\begin{figure}[htbp]
\centerline{\includegraphics[width=3in]{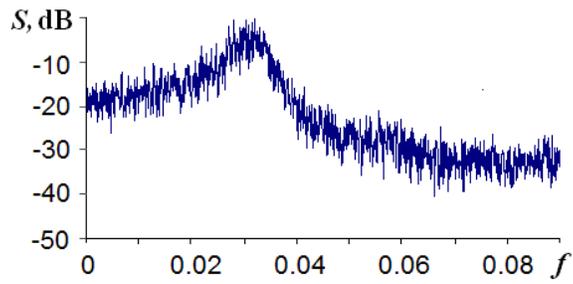}}
\label{fig10}
\caption{ The power spectrum of the signal generated by the system (\ref{eq16}) for $\nu $=3 and $\mu $=0.02.}
\end{figure}

\end{center}

Figure 11 shows histograms of the angles between stable and unstable
subspaces in the system (\ref{eq16}) obtained numerically for $\nu $=3 with $\mu
$=0.02 and 0.13. The distributions do not include angles close to zero, so,
the test confirms the hyperbolic nature of the attractor.

\begin{figure}[htbp]
\centerline{\includegraphics[width=5.5in]{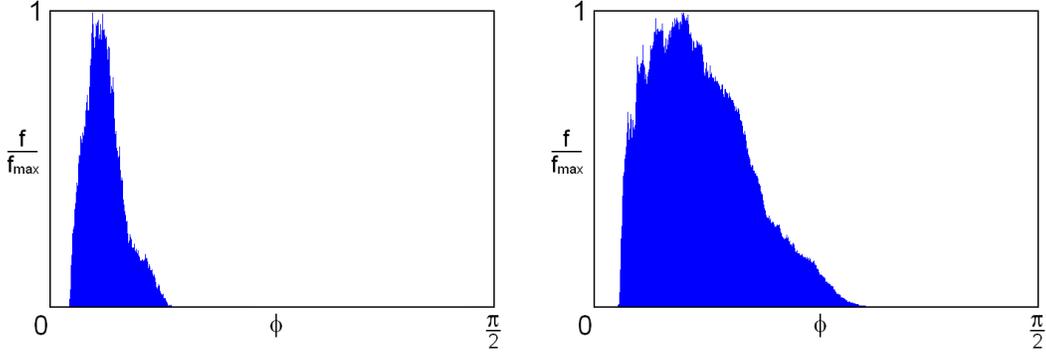}}
\label{fig11}
\caption{Verification of the criterion of angles in the linkage of three
self-rotators (\ref{eq16}) at $\mu $=0.02 and $\mu $=0.13 with $\nu $=3.}
\end{figure}

\section{Systems of three rotators with potential \\ interaction}

From systems with mechanical constraints let us move now to a class of systems
where the interaction of three constituting components (rotators) is
introduced by a potential function depending on the angle variables. We impose
a specific form of this function in such a way that the potential minimum is
reached on a surface just corresponding to that defined by the equation of
mechanical constraint in the previous examples:
$U(\theta _1 ,\theta _2 ,\theta _3 ) = \textstyle{1 \over 2}(\cos \theta _1 + \cos \theta _2 + \cos \theta _3 )^2$. 
Instead of equations (\ref{eq3}) we write now
\begin{equation}
\label{eq17}
\ddot {\theta }_i = - \partial U / \partial \theta _i + M_i = (\cos \theta
_1 + \cos \theta _2 + \cos \theta _3 )\sin \theta _1 + M_i ,\,\,i = 1,2,3,
\end{equation}
where $M_{1,2,3}$ designate external torques supplied to the rotators.

We will consider two options.

In the first case, we assume that the applied torques are defined according
to (\ref{eq10}) and ensure that the system tends to a level of constant kinetic energy.
The set of equations reads
\begin{equation}
\label{eq18}
\ddot {\theta }_i = \nu [\mu - \textstyle{1 \over 2}(\dot {\theta }_1^2 +
\dot {\theta }_2^2 + \dot {\theta }_3^2 )]\dot {\theta }_i - (\cos \theta _1
+ \cos \theta _2 + \cos \theta _3 )\sin \theta _i ,\,\,i = 1,2,3.
\end{equation}
This system was proposed and particularly studied earlier in Ref.~\cite{29}.

In the second case, we define the torques by (\ref{eq15}), so the system is composed of three self-rotators interacting due to the potential forces, and the
equations look as follows:
\begin{equation}
\label{eq19}
\ddot{\theta}_i=\nu(\mu-\textstyle{1\over2}\dot{\theta}_i^2)\dot{\theta}_i-(\cos\theta_1+\cos\theta_2+\cos\theta_3)\sin\theta_i,\,\,i=1,2,3.
\end{equation}

Figure 12 illustrates the transient processes in the systems (\ref{eq18}) and (\ref{eq19})
starting from a state of low energy. In both cases, the result of the
transient process is a chaotic self-oscillating mode where the kinetic energy
irregularly fluctuates around a mean value. These fluctuations are small in
amplitude and have relatively short time scale in the first model. In the
second model, the magnitude of the energy undulation is much larger, and the
time scale is notably larger: the energy dependence is similar to that for a
model with a mechanical constraint in Figure 7.

\begin{figure}[htbp]
\centerline{\includegraphics[width=5.5in]{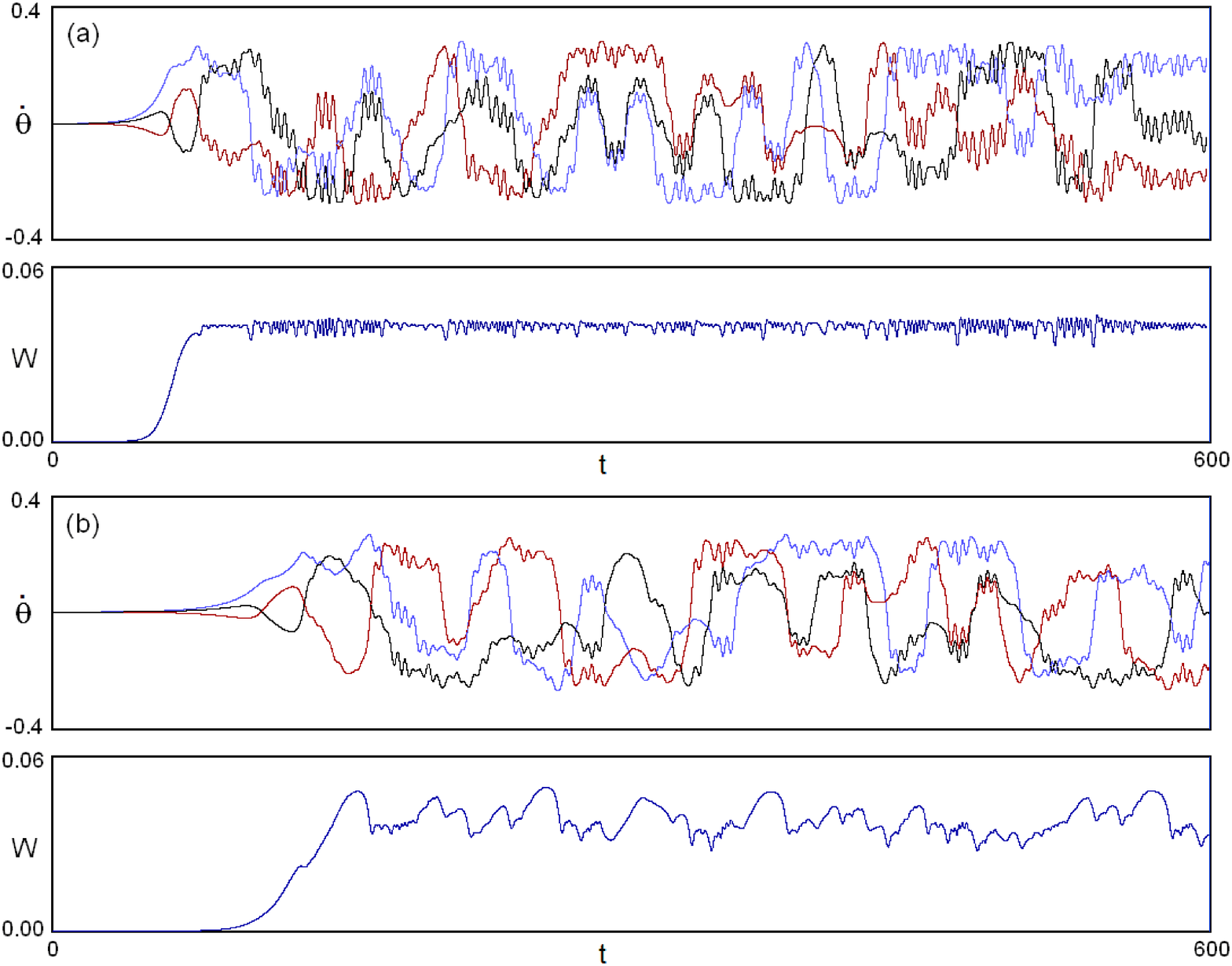}}
\label{fig12}
\caption{Generalized velocities and kinetic energy versus time in the
transient process for system (\ref{eq18}) at $\mu $=0.04, $\nu $=3 (a) and in
system (\ref{eq19}) at $\mu $=0.02, $\nu $=3 (b).}
\end{figure}

Figure 13 shows trajectories in the configuration space for the systems (\ref{eq18})
and (\ref{eq19}). For small values of the parameter $\mu $, which correspond to
small average energy of the sustained self-oscillatory mode, the orbits are
close to the two-dimensional surface defined by the equation 
$\cos \theta _1 + \cos \theta _2 + \cos \theta _3 = 0$, which is consistent with the
mechanical constraint imposed in models of two previous sections (diagram (a)
and (c)). Hence, we guess that the hyperbolic nature of the dynamics in this
domain yet persists. However, one can observe that the trajectory is
"fluffed up" transversally to the surface that reflects the occurrence of
fluctuations in the potential energy in the course of the motion in the
sustained mode. This effect is still negligible for small $\mu $, but it
becomes more pronounced with increase of the parameter, as seen in
diagrams (b) and (d). We expect (and this is supported by numerical
calculations, see below) that due to this effect the dynamics may change its
nature, and ceases to be hyperbolic.

Figure 14 demonstrates the dependence of the average kinetic energy
oscillations and standard deviation of the energy on parameter $\mu $ for
the models (\ref{eq18}) and (\ref{eq19}). As is
seen, in the second case the energy fluctuations are much more pronounced.
This is not surprising, since the models (\ref{eq18}) and (\ref{eq19}) are modifications of
the models (\ref{eq13}) and (\ref{eq16}) with potential interaction instead of the
mechanical constraint, and in the model (\ref{eq13}) the energy fluctuations are
excluded altogether. In diagram (b), approximately at $\mu  \approx $0.18, one
observes a sharp change in the dynamical behavior: at larger $\mu $ a regular self-oscillation 
mode occurs instead of chaos; this mode corresponds to an
attractive limit cycle in the phase space.

\begin{figure}[htbp]
\centerline{\includegraphics[width=5.5in]{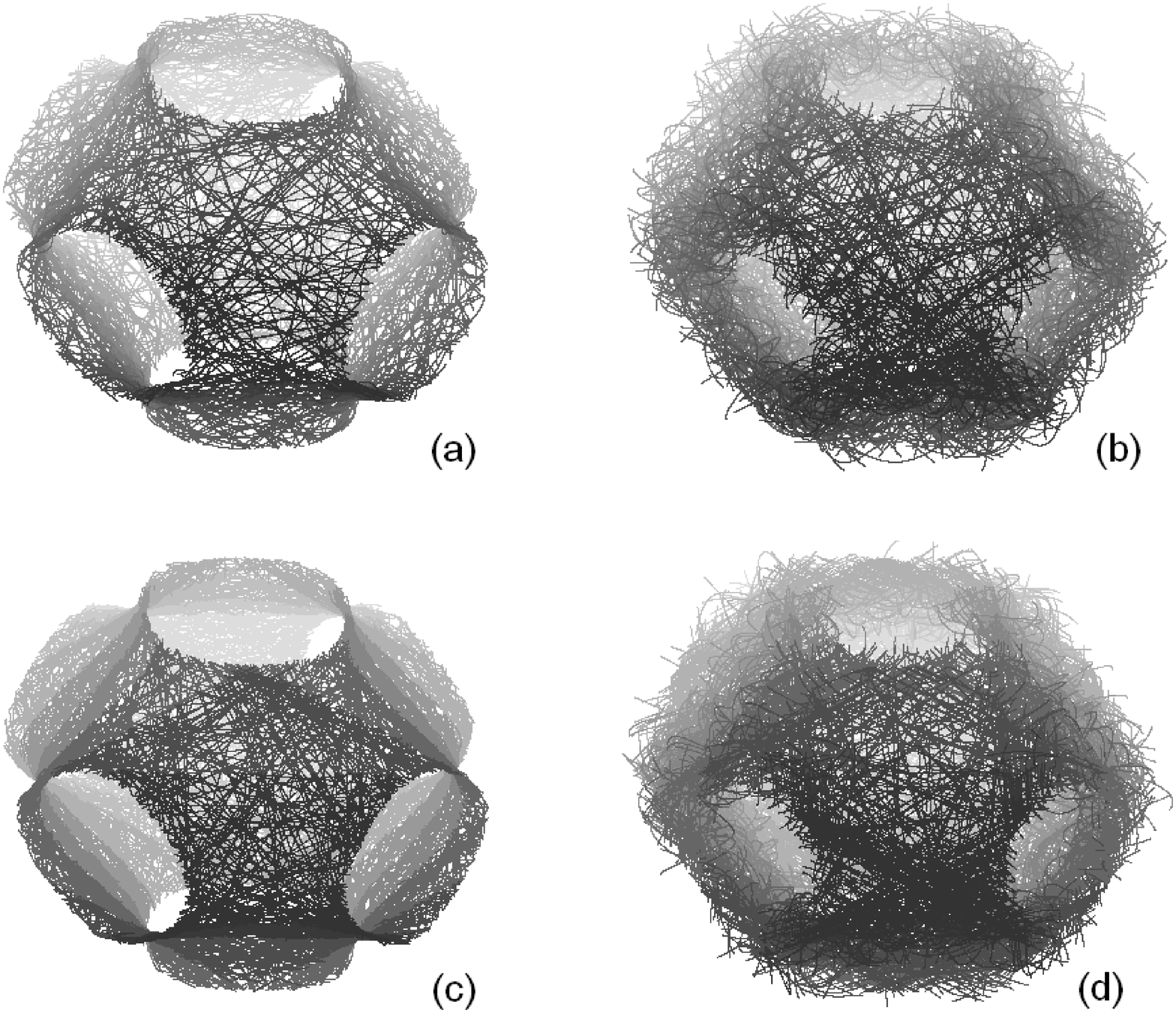}}
\label{fig13}
\caption{ Trajectories in the three-dimensional configuration space on
attractors of the system (\ref{eq18}) at $\mu $ = 0.04 (a) and $\mu $ = 0.25 (b), and of the system (\ref{eq19}) at $\mu $ = 0.02 (c) and $\mu $ = 0.13 (d) for $\nu=3$.}
\end{figure}

\begin{figure}[htbp]
\centerline{\includegraphics[width=5.4in]{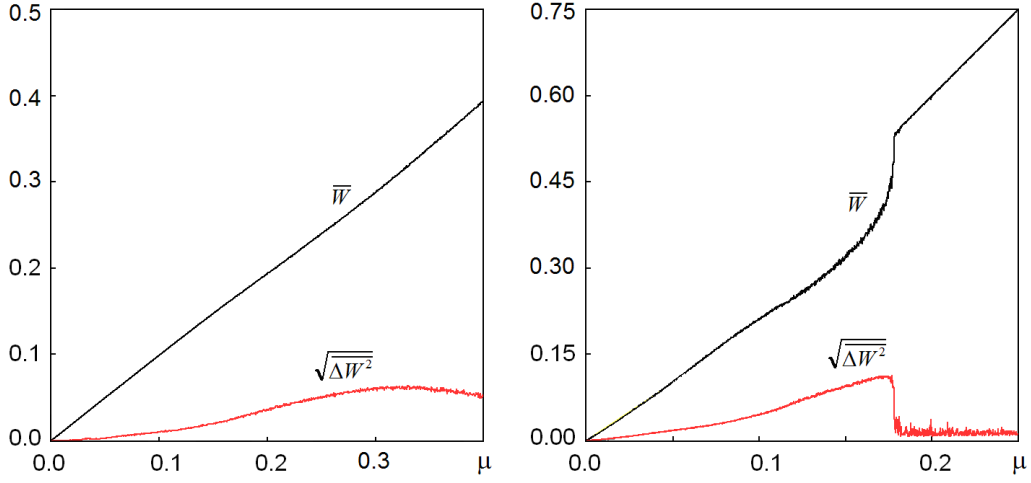}}
\label{fig14}
\caption{ The average kinetic energy of oscillations and its standard
deviation depending on parameter $\mu $ for three rotators with potential
interaction as obtained numerically for the models (\ref{eq18}) (a) and (\ref{eq19}) (b) at
$\nu $ = 3.}
\end{figure}
\begin{figure}[htbp]
\centerline{\includegraphics[width=3.5in]{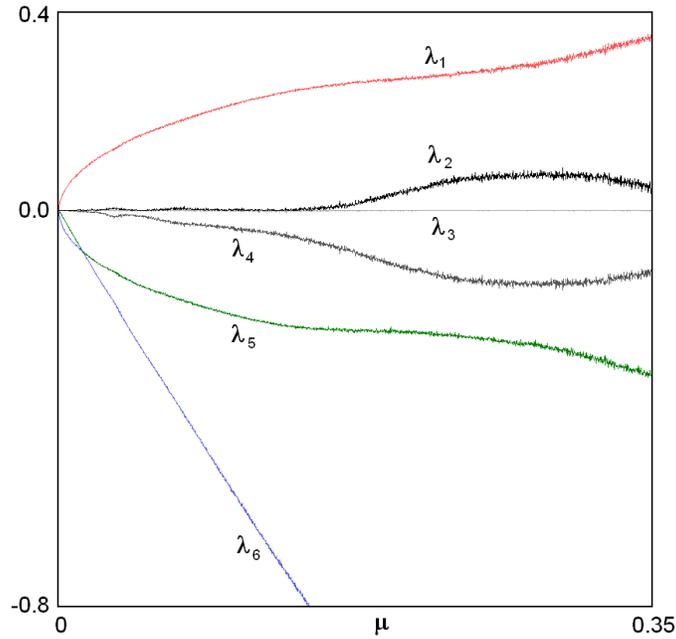}}
\label{fig15}
\caption{ Lyapunov exponents of the system (\ref{eq18}) composed of three rotators with interaction potential depending on the parameter $\mu $ at $\nu $=3.}
\end{figure}

Figures 15 and 16 show graphs of Lyapunov exponents depending on the
parameter $\mu $ for the models (\ref{eq18}) and (\ref{eq19}) calculated with the Benettin
algorithm. For these systems, we must take into account all six exponents as
there is no reason to exclude any of them from consideration. The
presence of a zero exponent is explained by the autonomous nature of the
system; it is associated with the perturbation vector tangent to the reference
trajectory.

For the system (\ref{eq18}), throughout the parameter range shown in the picture, the
senior Lyapunov exponent is positive, indicating a chaotic nature of the
dynamics. Its parameter dependence is smooth, without drops accompanying
the appearance of "windows of regularity" in many systems with nonhyperbolic
attractors. The second exponent in the region of small $\mu $ is close to
zero, but then becomes positive. So for $\mu > 0.17$ we have two positive
exponents; this regime is referred to as hyperchaos~\cite{38}. Three exponents
remain negative in the whole range, although one of them is close to zero in
the region of small $\mu $. Obviously, at small $\mu $, where three
exponents are close to zero, we can suggest a possibility of approximate
description of the dynamics with replacement of the potential interaction by
the mechanical constraint with reduction to the model, discussed in Section
2. Also the possibility of approximation of the exponents $\lambda _{1}$ and
$\lambda _{5}$ by $\pm \mbox{const} \cdot \sqrt W $, as in the reduced
model, is evident from Figure 15.

For the system (\ref{eq19}) in the region of small $\mu $ we have one positive, one
zero, and four negative Lyapunov exponents. The dependence of the exponents
on the parameter is smooth here, without irregularities, which suggests that
the hyperbolic chaos persists, as in the reduced model of Section 3. The
senior Lyapunov exponent remains positive, and the dynamics is chaotic up to
$\mu  \approx $0.18. When this point is approached, brokenness arises and
progresses in the graph of the senior exponent, which apparently indicates
destruction of the hyperbolicity, though no visible drops to zero with
formation of regularity windows are distinguished. Then the chaos
disappears sharply, and the system manifests transition to the regular mode,
where the senior exponent is zero, and the others are negative, which
corresponds to the limit cycle.

Figure 17 shows power spectra of the signal $u(t) = \cos \theta _1 (t)$
generated by the systems (\ref{eq18}) and (\ref{eq19}) at relatively low values of $\mu $,
which correspond presumably to the hyperbolic chaos. They can be compared
with that shown in Figure 5 for the model (\ref{eq13}), certainly relating to the
hyperbolic chaos. The spectra are continuous, which corresponds to the chaotic
nature of the dynamics, and evidently are characterized by a relatively high
degree of uniformity.

\begin{figure}[htbp]
\centerline{\includegraphics[width=3.5in]{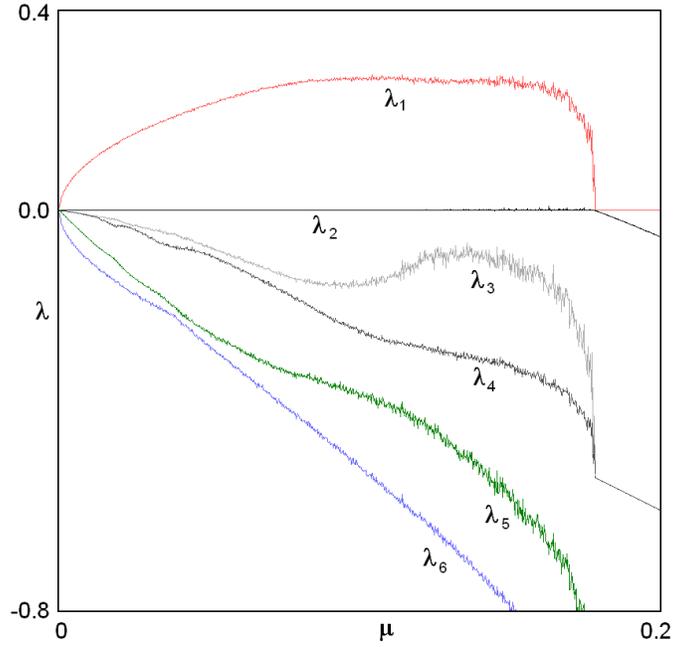}}
\label{fig16}
\caption{ Lyapunov exponents of the system (\ref{eq19}) composed of three
self-rotators with interaction potential, depending on the supercriticality
parameter $\mu $ with $\nu $=3.}
\end{figure}
\begin{figure}[htbp]
\centerline{\includegraphics[width=5.5in]{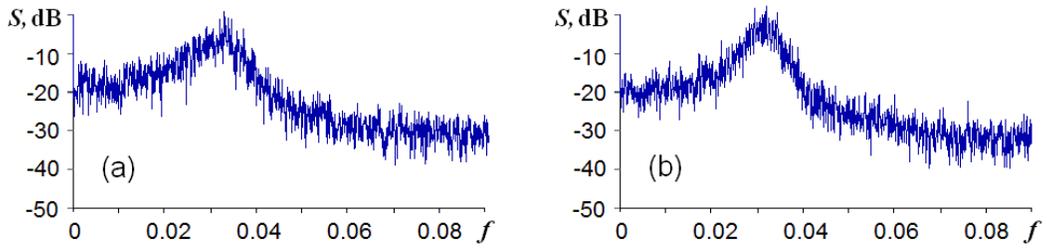}}
\label{fig17}
\caption{ Power spectra of the signal $u(t) = \cos \theta _1 (t)$ for $\nu
$=3 in the system (\ref{eq18}) at $\mu $=0.04 (a) and in the system (\ref{eq19}) $\mu $=0.02 (b).}
\end{figure}
\begin{figure}[htbp]
\centerline{\includegraphics[width=5in]{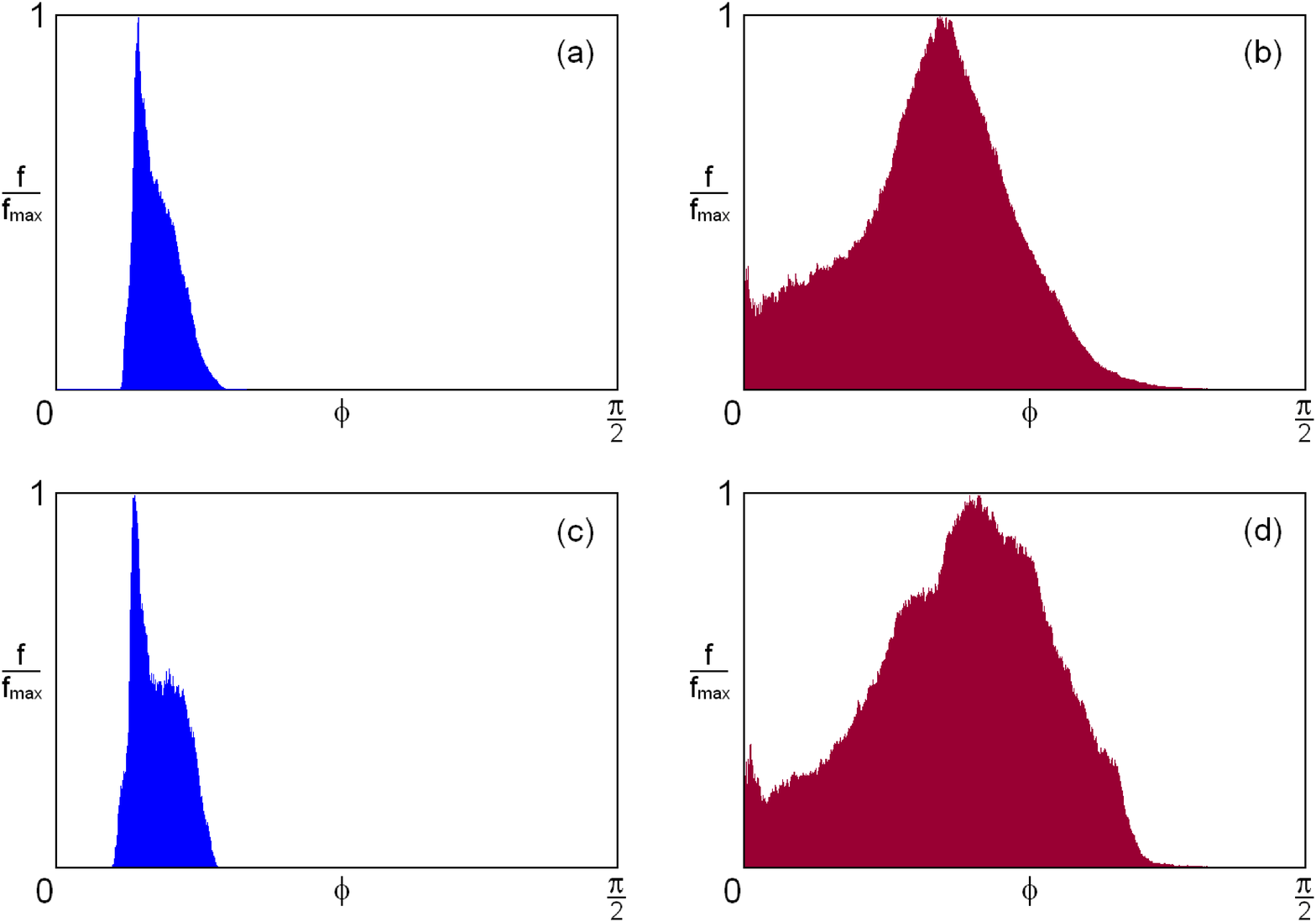}}
\label{fig18}
\caption{ Verification of the criterion of angles for systems of three
rotators with potential interaction in the model (\ref{eq18}) at $\mu $=0.04 (a) and $\mu $=0.25 (b), and in the model (\ref{eq19}) at $\mu $=0.02 (c) and $\mu $=0.13 (d). Histograms (a) and (c) show distributions separated
from zero, which indicates the hyperbolic nature of the attractor.
Diagrams (b) and (d) show the distributions with positive probability of
zero angles indicating occurrences of tangencies of stable and
unstable manifolds, which means violation of the hyperbolicity.}
\end{figure}

Figure 18 shows results of testing the attractors of models (\ref{eq18}) and (\ref{eq19}) using
the criterion of angles. Here the numerically obtained histograms are
plotted for the angles between stable and unstable subspaces of perturbation
vectors for typical chaotic phase trajectories. At small values of $\mu $
the distributions are disposed at some finite distance from zero,
i.e., the test confirms the hyperbolicity. This feature, however, is violated
somewhere in the course of increase of $\mu $. Observe that histograms
(b) and (d) clearly demonstrate the presence of angles close to zero, which
indicates the occurrence of tangencies of stable and unstable manifolds, and,
hence, signalizes about non-hyperbolic nature of the attractor. As this
phenomenon was not observed in the reduced models of Sections 2 and 3, it is
natural to assume that the destruction of hyperbolicity is linked with excursions of
phase trajectories relating to the attractor outside a narrow neighborhood
of the surface of equal potential in the configuration space.

\section{Conclusion}

Chaos associated with hyperbolic attractors in dissipative systems and with
Anosov dynamics in conservative systems is characterized by roughness, or
structural stability as a mathematically rigorously proven attribute.
Therefore, devices that generate such chaos should be preferred for any
practical applications of chaos as they are robust, i.e., they have low
sensitivity to various imperfections, variations of parameters, noise, etc.

A possible productive approach to design systems with hyperbolic chaos
starting from some abstract and artificial example of such dynamical
behavior consists in its modification to a physically realizable device by
variation of functions and parameters involved in the original equations.
While the variations are small enough, the hyperbolic nature of the dynamics
persists by virtue of the structural stability. However, if we depart
further and further from the original system, the hyperbolicity may break
up; its preservation must be monitored using some mathematically justified
quantitative criteria. One of them is based on analysis of statistical
distribution of angles between stable and unstable manifolds of relevant
phase trajectories to demonstrate absence of tangencies between these
manifolds.

In this article we have discussed several chaotic systems based on the
triple linkage hinge mechanism of Thurston, Weeks, Hunt, and MacKay. The
original system is composed of three elements able to rotate about fixed
axles with mechanical constraint imposed with some hinges and rods. We modify
the model and introduce activity and dissipation by adding appropriate terms
in the equations to get examples of systems with hyperbolic chaotic
attractors. In one version, these additional terms are chosen to provide a
tendency to approach a constant level of kinetic energy. In another more
pragmatic setup, each of the three basic elements is represented by a
self-rotator that is a subsystem, which, being isolated, tends to rotate in
one or other direction with constant angular velocity.

Results of numerical simulations are presented, and characteristics of the
attractors (Lyapunov exponents, power spectra) are discussed. Robust chaos
is observed in the models under study in wide ranges, being not destroyed
under variation of parameters. The calculations have been performed for
verification of the criterion of angles, and confirmation of the hyperbolic
nature of attractors is provided in the considered models, at least, for
self-oscillatory modes of small supercriticality.

An alternative approach to a more rigorous substantiation of hyperbolicity may
be based on computer verification of the so-called Alekseev cone criterion~\cite{39,4,40,13}.
In Ref.~\cite{28a} the authors consider the application of the cone criterion to justification of Anosov dynamics for
linkages where the curvature is not globally negative (some areas of positive curvature are present). 
One more approach consists in analyzing invariant measure; on hyperbolic attractors it must correspond to the concept of Sinai -- Ruelle -- Bowen measures~\cite{3,4,13}.

Besides the models with mechanical constraints, we have suggested and
considered systems manifesting analogous dynamical behavior, where
interaction of the rotators is provided by potential forces. In principle,
it opens up a prospect to go away from mechanics and
implement systems of a different nature, e.g., electronic devices operating as
generators of robust chaos. They will manifest hyperbolic dynamics, roughly
uniform in continuous time, in contrast to previously proposed systems~\cite{13,14}; apparently, 
it will improve essentially the spectral properties of the
generated chaotic signals.

\bigskip
\textit{This work was supported by RNF grant No 15-12-20035.}


\newpage
\begin{thebibliography}{99}
\bibitem{1}
Smale, S., Differentiable Dynamical Systems, Bull. Amer. Math. Soc. (NS),
1967, vol. 73, pp. 747-817.
\bibitem{2}
Shilnikov, L., Mathematical Problems of Nonlinear Dynamics: A Tutorial,
Internat. J. Bifur. Chaos Appl. Sci. Engrg., 1997, vol. 7, no. 9, pp.
1953-2001.
\bibitem{3}
Dynamical Systems 9: Dynamical Systems with Hyperbolic Behaviour, D.V.Anosov
(Ed.), Encyclopaedia Math. Sci., vol. 9, Berlin: Springer, 1995.
\bibitem{4}
Katok, A. and Hasselblatt, B., Introduction to the Modern Theory of
Dynamical Systems, Cambridge: Cambridge University Press, 1996.
\bibitem{5}
Afraimovich, V. and Hsu, S.-B., Lectures on Chaotic Dynamical Systems,
AMS/IP Studies in Advanced Mathematics, vol. 28, Providence, RI: Amer. Math.
Soc., Somerville, MA: International Press, 2003.
\bibitem{6}
Pesin, Ya.B., Lectures on partial hyperbolicity and stable ergodicity.
Zurich lectures in advanced mathematics, European Mathematical Society,
2004.
\bibitem{7}
Bonatti, C., Diaz, L.J., Viana, M., Dynamics Beyond Uniform Hyperbolicity. A
Global Geometric and Probobalistic Perspective. Encyclopedia of Mathematical
Sciences, vol.102, Berlin, Heidelberg, New-York: Springer, 2005.
\bibitem{8}
Kuznetsov, S.P., Example of a Physical System with a Hyperbolic Attractor of
the Smale--Williams Type, Phys. Rev. Lett., 2005, vol. 95, 144101.
\bibitem{9}
Kuznetsov, S.P. and Seleznev, E.P., Strange Attractor of Smale--Williams
Type in the Chaotic Dynamics of a Physical System, Zh. Eksper. Teoret. Fiz.,
2006, vol. 129, no. 2, pp. 400-412 [J. Exp. Theor. Phys., 2006, vol. 102,
no. 2, pp. 355-364].
\bibitem{10}
Isaeva, O.B., Jalnine, A.Y., Kuznetsov, S.P., Arnold's cat map dynamics in a
system of coupled nonautonomous van der Pol oscillators, Physical Review E,
2006, vol. 74, no 4, p. 046207.
\bibitem{11}
Kuznetsov, S.P. and Pikovsky, A. Autonomous Coupled Oscillators with
Hyperbolic Strange Attractors, Physica D, 2007, vol. 232, pp. 87-102.
\bibitem{12}
Kuznetsov, S.P., Example of Blue Sky Catastrophe Accompanied by a Birth of
Smale--Williams Attractor, Regular and Chaotic Dynamics, 2010, vol. 15, nos.
2-3, pp. 348-353.
\bibitem{13}
Kuznetsov, S.P., Hyperbolic Chaos: A Physicist's View, Berlin: Springer,
2012.
\bibitem{14}
Kuznetsov, S.P., Dynamical Chaos and Uniformly Hyperbolic Attractors: From
Mathematics to Physics, Phys. Uspekhi, 2011, vol. 54, no. 2, pp. 119-144;
see also: Uspekhi Fiz. Nauk, 2011, vol. 181, pp. 121-149.
\bibitem{15}
Kuznetsov, S.P., Plykin type attractor in electronic device simulated in
MULTISIM, Chaos: An Interdiskiplinary Journal of Nonlinear Science, 2011,
vol. 21, no 4, p. 043105.
\bibitem{16}
Isaeva, O.B., Kuznetsov, S.P., and Mosekilde, E., Hyperbolic chaotic
attractor in amplitude dynamics of coupled self-oscillators with periodic
parameter modulation, Physical Review E., 2011, vol. 84, no. 1, p. 016228.
\bibitem{17}
Isaeva, O.B., Kuznetsov, A.S., Kuznetsov, S.P., Hyperbolic chaos in
parametric oscillations of a string, Rus. J. Nonlin. Dyn., 2013, vol. 9, no.
1, pp. 3-10
\bibitem{18}
Kuznetsov S.P., Kuznetsov A.S., Kruglov V.P., Hyperbolic chaos in systems
with parametrically excited patterns of standing waves, Rus. J. Nonlin.
Dyn., 2014, vol. 10, no.~3, pp. 265-277. (Russian.)
\bibitem{19}
Jalnine, A.Y., Hyperbolic and non-hyperbolic chaos in a pair of coupled
alternately excited FitzHugh--Nagumo systems, Communications in Nonlinear
Science and Numerical Simulation, 2015, vol. 23, no. 1, pp. 202-208.
\bibitem{20}
Kuznetsov, S.P., Some mechanical systems manifesting robust chaos, Nonlinear
Dynamics and Mobile Robotics, 2013, vol. 1, no. 1, pp. 3-22.
\bibitem{21}
Borisov, A.V., Kazakov, A.O., and Kuznetsov, S.P., Nonlinear dynamics of the
rattleback: a nonholonomic model, Physics-Uspekhi, 2014, vol. 57, no. 5, pp.
453-460; see also: Uspekhi Fiz. Nauk, 2014, vol. 184, pp. 493-500.
\bibitem{22}
Kuznetsov, S.P., Plate falling in a fluid: Regular and chaotic dynamics of
finite-dimensional models, Regular and Chaotic Dynamics, 2015, vol. 20, no.
3, pp. 345-382.
\bibitem{23}
Borisov, A.V., Mamaev, I.S., On the motion of a heavy rigid body in an ideal
fluid with circulation, Chaos: An Interdiskiplinary Journal of Nonlinear
Science, 2006, vol. 16, no. 1, p. 013118.
\bibitem{24}
Borisov, A.V., Jalnine, A.Y., Kuznetsov, S.P., Sataev, I.R., and Sedova,
J.V., Dynamical phenomena occurring due to phase volume compression in
nonholonomic model of the rattleback, Regular and Chaotic Dynamics, 2012,
vol. 17, no. 6, pp. 512-532.
\bibitem{25}
Thurston, W.P. and Weeks, J.R., The mathematics of three-dimensional
manifolds, Sci. Am., 1984, vol. 251, pp. 94-106.
\bibitem{25a}
Kozlov, V.V., Topological obstacles to the integrability of natural mechanical systems, Soviet Math. Dokl., 1980, vol. 20, pp. 1413-1415.
\bibitem{27}
Anosov, D.V., Geodesic flows on closed Riemannian manifolds of negative
curvature, Trudy Mat. Inst. Steklov, 1967, vol. 90, pp. 3-210.
\bibitem{28}
Balazs, N.L. and Voros, A., Chaos on the pseudosphere, Physics Reports,
1986, vol. 143, no. 3, pp. 109-240.
\bibitem{26}
Hunt, T.J. and MacKay, R.S., Anosov Parameter Values for the Triple Linkage
and a Physical System with a Uniformly Chaotic Attractor, Nonlinearity,
2003, vol. 16, pp. 1499-1510.
\bibitem{28a}
Magalh\~{a}es, M. L. S. and Pollicott, M., Geometry and dynamics of planar
linkages, Communications in Mathematical Physics, 2013, vol. 317, no. 3,
pp. 615-634.
\bibitem{28b}
Kourganoff, M., Anosov geodesic flows, billiards and linkages,
arXiv:1503.04305, 2015, pp. 1-27.
\bibitem{28c}
Kozlov, V.V., Closed orbits and chaotic dynamics of a charged particle in a periodic electromagnetic field, Regular and Chaotic Dynamics, 1997, vol. 2,
no. 1, pp. 3-12.
\bibitem{29}
Kuznetsov, S.P., Chaos in the system of three coupled rotators: from Anosov
dynamics to hyperbolic attractor, Proceedings of Saratov University -- New
series. Series Physics, 2015, vol. 15, no. 2, pp. 5-17. (Russian.)
\bibitem{30}
Lai, Y.-C., Grebogi, C., Yorke, J. A., Kan, I., How often are chaotic
saddles nonhyperbolic? Nonlinearity, 1993, vol. 6, pp. 779-798.
\bibitem{31}
Anishchenko, V.S., Kopeikin, A.S., Kurths, J., Vadivasova, T.E., Strelkova
G.I., Studying hyperbolicity in chaotic systems, Physics Letters A, 2000,
vol. 270, pp. 301-307.
\bibitem{32}
Ginelli, F., Poggi, P., Turchi, A., Chat\'{e}, H., Livi, R., Politi, A.,
Characterizing Dynamics with Covariant Lyapunov Vectors, Phys. Rev. Lett.,
2007, vol. 99, p. 130601.
\bibitem{33}
Kuptsov, P.V., Fast numerical test of hyperbolic chaos, Physical Review E,
2012, vol. 85, no 1, p. 015203.
\bibitem{34}
Gantmakher, F.R., Lectures in analytical mechanics, Moscow: Mir Publishers,
1970.
\bibitem{35}
Goldstein, H., Poole, C.P., Safko, J.L., Classical Mechanics (3rd ed.).
Addison-Wesley, 2001.
\bibitem{36}
Benettin, G., Galgani, L., Giorgilli, A., and Strelcyn, J.-M., Lyapunov
Characteristic Exponents for Smooth Dynamical Systems and for Hamiltonian
Systems: A Method for Computing All of Them, Meccanica, 1980, vol. 15, pp.
9-30.
\bibitem{37}
Jenkins, G.M. and Watts, D.G.: Spectral analysis and its application. San
Francisco: Holden-Day, Inc., 1968.
\bibitem{38}
R\"{o}ssler, O. E., An equation for hyperchaos, Physics Letters A, 1979, vol.
71, no. 2, pp.~155-157.
\bibitem{39}
Sinai, Y.G., Stochasticity of dynamical systems. In: Gaponov-Grekhov, A.V.
(ed.), Nonlinear Waves, Moscow: Nauka, 1979, pp.192-212. (Russian.)
\bibitem{40}
Kuznetsov, S.P. and Sataev, I.R., Hyperbolic attractor in a system of
coupled non-autonomous van der Pol oscillators: Numerical test for expanding
and contracting cones, Physics Letters A, 2007, 365, nos. 1-2, pp. 97-104.
\end{thebibliography}
\end{document}